\documentclass[10pt,letterpaper,twocolumn]{article} 

\usepackage{ol2}
\usepackage[draft]{hyperref}
\usepackage{amsmath}

\begin{document}

\twocolumn[ 

\title{Multifrequency broadband tapered plasmonic nanoantennas}


\author{I. S. Maksymov,$^{1,*}$ A. R. Davoyan,$^1$ C. Simovski,$^{2,3}$ P. Belov$^2$ and Yu. S. Kivshar$^{1,3}$}

\address{
$^1$Nonlinear Physics Centre, Australian National University, Canberra ACT 0200, Australia \\
$^2$ Department of Radio Science and Engineering, Aalto University, Aalto 00076, Finland \\
$^3$ National Research University of Information Technologies, Mechanics, and Optics (ITMO), St Petersburg 197101, Russia \\
$^*$Corresponding author: mis124@physics.anu.edu.au
}

\begin{abstract} We suggest a novel multifrequency broadband plasmonic Yagi-Uda-type nanoantenna equipped with an array of tapered directors. Each director can be used for the excitation of the antenna by nanoemitters matched spectrally with the director resonant frequency and placed in the director near-field region. Multifrequency operation of nanoantennas provides tremendous opportunities for broadband emission enhancement, spectroscopy and sensing. By the principle of reciprocity, the same tapered nanoantenna architecture can be used both as a transmitter and/or as a receiver, thus being useful for creating a broadband wireless communication system.\end{abstract}


 ] 

\noindent Emission of nanoscale sources such as e.g. quantum dots \cite{shields} is strongly enhanced and highly directed if the the emitter is placed in the close vicinity to the feed of a plasmonic Yagi-Uda nanoantenna \cite{curto, kosako, dregely}. Originally proposed for radio-frequency (RF) applications \cite{balanis}, Yagi-Uda antennas, downscaled to nanometric dimensions and operating at visible or telecom wavelengths, offer significant advantages over other competing architectures for optical nanoantennas (see e.g. Ref. \cite{novotny} for a review). The advantages are based on RF antenna design principles relying on careful selection of the length, dimensions and spacing of the antenna elements \cite{balanis}. 

Recently, it was shown that a proper design of directors of a plasmonic Yagi-Uda nanoantenna can sharpen up the angular emission diagram and increase the antenna gain \cite{arxiv}. The suggested antenna architecture also allows shrinking the antenna longitudinal dimension by decreasing tenfold the spacing between the antenna elements as compared with the inter-element spacings in classically designed Yagi-Uda nanoantennas \cite{curto, kosako}. 

All Yagi-Uda nanoantennas demonstrated so far \cite{curto, kosako, dregely, arxiv} are single-frequency and narrowband due to the wavelength selectivity of the director arrays. However, it is very important for communication systems to develop unidirectional broadband nanoantennas for the telecom transmission band spanning between $1.3 \mu$m and $1.55 \mu$m \cite{gisin, shields}. While nanoemitters operating at telecom wavelengths have been achieved and demonstrated \cite{shields}, multifrequency broadband nanoantennas capable of enhancing the emission from one or several local nanoemitters have not yet been explored. 

In this Letter, we suggest and study theoretically a multifrequency broadband plasmonic Yagi-Uda nanoantenna consisting of a reflector and directors gradually tapered towards the end of the antenna. Each of the antenna directors can perform as an active element driven by nanoemitters placed in the near-field. As the optical response of the antenna depends on the size of the excited element, the excitation of shorter antenna elements will shift the antenna operation towards shorter wavelengths. Given that the length of directors is gradually tapered, a smooth operating wavelength blue-shifting is expected with the same antenna architecture excited by properly choosing the position and the wavelength of the emitter. We demonstrate that an overlap of multiple resonances supported by the antenna results in a broad operating band with a nearly constant spontaneous emission enhancement. 

A somewhat similar idea of tapered metamaterials has recently been suggested in \cite{luukkonen}. Also, plasmonic gratings capable of focusing light into subwavelength focal points at \emph{several} wavelengths have been demonstrated in Ref. \cite{boriskina}. However, the former structure is designed to offer a broadband radiation suppression, while the latter is driven from the far-field and cannot be coupled to nanoemitters. The architecture suggested in this Letter meets a lack of nanoemitter-driven antennas supporting tens of operating wavelengths. It offers considerable advantages in pursuing analogy with highly optimized RF antennas \cite{novotny} and provides tremendous opportunities for broadband optical manipulation \cite{grigorenko}, nonclassical light emission \cite{maksymov}, high-resolution spectroscopy \cite{schuck}, etc. Employed as a transmitter and a receiver, the tapered nanoantenna also opens up ways for multiband wireless communication systems \cite{gisin}.

\begin{figure}[htb]
\centerline{\includegraphics[width=8.3cm]{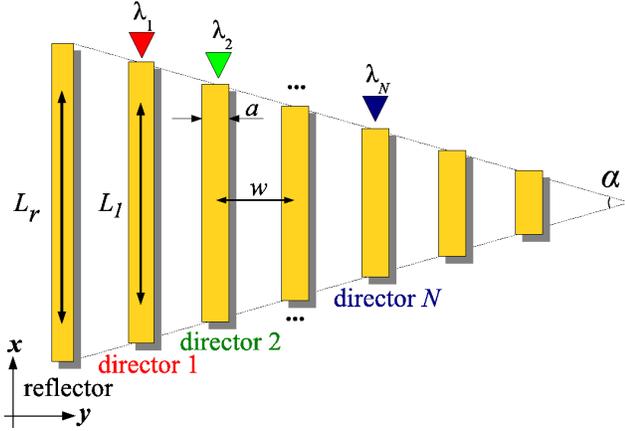}}
\caption{ (color online) Schematic view of a plasmonic Yagi-Uda nanoantenna consisting of a reflector of length $L_{r}$ and directors tapered according to the tapering angle $\alpha$ starting from the length of the first director $L_{1}$. The triangles denote the position of nanoemitters exciting the antenna via the corresponding antenna directors. The emitter wavelengths are $\lambda_{1}>\lambda_{2}>...>\lambda_{N}$.}
\end{figure}

We consider a silver plasmonic Yagi-Uda antenna (Fig. $1$) consisting of a reflector and equally spaced directors. We follow \cite{arxiv} and introduce tapering of the antenna directors in order to decrease the spacing between the antenna elements tenfold also enhancing the emission directivity offered by the antenna. In order to design the length of the tapered elements, we use the effective wavelength rescaling approach \cite{novotny1} introduced to account for inevitable losses in silver at telecom wavelengths \cite{palik}. We study the directivity and the beamwidth of the antennas with different tapering angles. We use a commercial finite-element (FEM) COMSOL Multiphysics software to calculate the far-field emission diagrams as \cite{balanis}
\begin{eqnarray}
D(\theta)=\frac{2{\pi}P(\theta)}{P_{\rm{loss}}+P_{\rm{rad}}}
\label{eq:one},
\end{eqnarray}
where $P(\theta)$ is the angular radiated power, $P_{\rm{rad}}$ denotes the Poynting vector flux through a closed contour in the far-field region and $P_{\rm{loss}}$ is the absorption losses in metal elements of the antenna. 

We find a pronounced maximum of the directivity and a minimum of the beamwidth at $\alpha=6.6^{o}$ at the principal design wavelength $1.5 \mu$m \cite{arxiv}. By choosing the width of the reflector and all directors as $a=50$ nm and the spacing between all elements as $w=30$ nm, we obtain the length of the reflector $L_{r}=485$ nm and the length of the 1st director (adjacent to the reflector) $L_{1}=390$ nm. As the impact of the tapering can be observed only if the antenna is sufficiently long, we equip the antenna with 41 tapered directors, thus offering enough length to satisfy the tapering angle.

Hereafter, we demonstrate that the antenna architecture with the aforementioned design parameters not only performs well at $1.5 \mu$m, but also is suitable for efficient operation at lower wavelengths from the telecom window. We use a finite-difference time-domain (FDTD) method \cite{taflove} to calculate the power radiated by the nanoantenna. We integrate the Poynting vector flux through a virtual plane of length $2 \mu$m placed $5 \mu$m from the last director of the antenna and orientated perpendicularly to the \textit{y}-coordinate direction (see Fig. $1$). A series of simulations is performed for the Yagi-Uda nanoantenna excited by turns via the first 10 director elements with a point broadband field source placed $5$ nm from the director's edge. The normalization of the antenna emission spectra to the free-space emission spectra results in the corresponding spontaneous emission enhancement spectra discussed hereafter.

\begin{figure}[htb]
\centerline{\includegraphics[width=8.3cm]{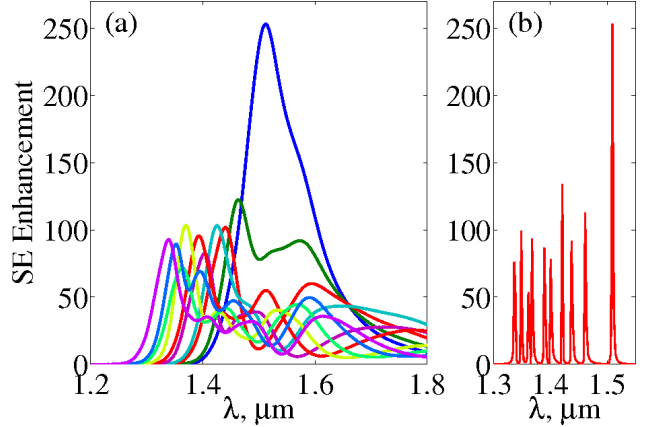}}
\caption{ (color online) (a) Spontaneous emission enhancement spectra of the nanoantenna excited with an \textit{x}-polarized emitter via different director elements. (b) Spontaneous emission enhancement spectra of the nanoantenna excited simultaneously by 10 \textit{x}-polarized nanoemitters.}
\end{figure}

The spontaneous emission enhancement spectra of the nanoantenna excited by \textit{x}-polarized emitters [Fig. $2$(a)] demonstrate a multifrequency operation of the nanoantenna. Placing the emitters near the antenna directors leads to the excitation of the director's fundamental dipole modes at corresponding resonant wavelengths.  The excitation of the fundamental mode of the longest antenna element with the resonant wavelength of about $1.5 \mu$m results in the antenna resonances and unidirectional emission pattern formation at nearly the same operating wavelength. The use of shorter antenna elements shifts the antenna operating wavelength towards shorter wavelengths. According to this physical picture, the maximum of the spontaneous emission enhancement is observed at $1510$ nm for the 1st director excited by the emitter and shits towards $1330$ nm for the 10th director, thus covering 10 transmission bands between $1.3 \mu$m and $1.5 \mu$m. Also, the striking feature is the broadband operation: the spontaneous emission spectra for different emitter positions overlap and the maximum amplitudes of the spectra peaks are nearly constant, except the rightmost peak in Fig. $2$(a). Qualitatively the same conclusions hold also for the antenna excited by different in-plane unpolarized emitters, evidencing that a one-by-one excitation of the tapered antenna presents a reliable solutions for both broadband and multifrequency operating regimes even if the orientation of the emitters cannot be fully controlled \cite{bleuse}.

Apart from the excitation according to the one-by-one scheme, the antenna can be also excited by an ensemble of the emitters placed in the near-field of the corresponding directors and emitting at the corresponding wavelengths. As shown in Fig. $2$(b), a collective excitation with narrow emission linewidth emitters converts the nanoantenna into an efficient multifrequency plasmon-enhanced light emitter. 

\begin{figure}[htb]
\centerline{\includegraphics[width=7.5cm]{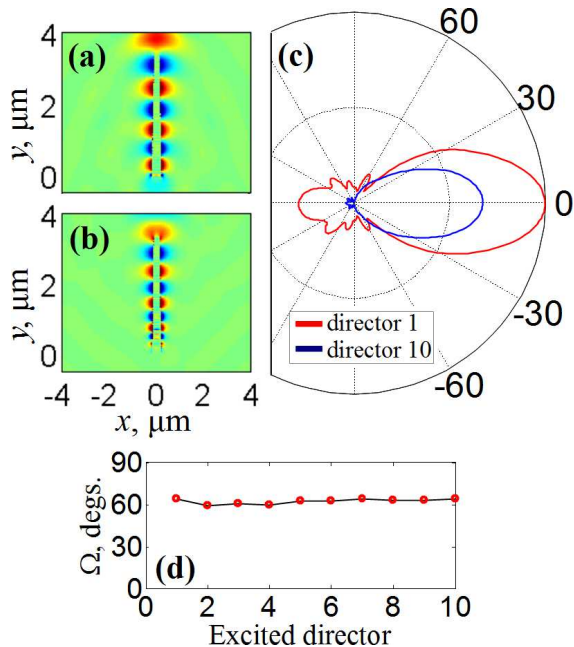}}
\caption{(color online) Near-field $E_x$ electric field profiles of the nanoantenna excited via the 1st (a) and 10th (b) directors. In all panels, the same antenna positions and the color-scale are used. (c) Far-field angular emission diagrams of the antenna excited via the 1st (red line) and 10th directors (blue line). (d) Beamwidth $\Omega$ of the antenna excited  by turns via the first ten directors. The black line is a guide to the eye.}
\end{figure}

Just for the illustration purpose, in Figs. $3$(a, b) we show the calculated near-field profiles for the antenna excited via the 1st and 10th directors. A close observation of the near-field patterns reveals that the excitation of the antenna via a shorter director element with a shorter-wavelength emitter results in a shorter effective length of the antenna. The antenna elements not used for directors form a multi-element reflector array, which is known to efficiently suppress the backward emission from the antenna \cite{balanis}. This effect is clearly seen by comparing the backward lobes of the antenna emission diagrams in Fig. $3$(c). Also, unlike at RF, at telecom wavelengths the antenna emission is strongly affected by a decrease in the metal losses due to the decrease in both the operating wavelength and the number of directors on the way of the end-fire beam produced by the antenna. The sum contribution of these effects maintains the beamwidth of the antenna \cite{balanis} at nearly the same level of about $65^{o}$ for all excited directors [Fig. $3$(d)], thus ensuring equal conditions for the collection of the emitted light within a certain cone given by the numerical aperture of the detection optics. By the principle of reciprocity, it also guarantees equal reception conditions with the same antenna employed as a receiver.   

In conclusion, we have investigated theoretically the physics of a multifrequency broadband tapered plasmonic Yagi-Uda nanoantenna that merits experimental investigation for broadband emission enhancement from nanoemitters and can be used both for receiver and transmitter in on-chip wireless communication systems. The bandwidth of the antenna can be geometrically tuned by changing the tapering angle and further extended up to the visible range by exciting the antenna by turns via the first 20, 25 and 30 and so on directors, respectively. 

This work was supported by the Australian Research Council (Australia) and the Ministry of Education and Science in Russia.


\pagebreak

\section*{Informational Fourth Page}

\end{document}